\documentclass[11pt]{article}
\usepackage{amssymb}
\usepackage{latexsym}
\begin{document}
\setlength{\baselineskip}{15pt}
\title{Generalization of the separation of variables in the \\ 
Jacobi identities for finite-dimensional Poisson systems}
\author{ \mbox{} \\ \mbox{} \\ Benito Hern\'{a}ndez-Bermejo $^1$}
\date{}

\maketitle
\begin{center}
{\em Departamento de F\'{\i}sica. Escuela Superior de Ciencias Experimentales y 
Tecnolog\'{\i}a. \\ 
Universidad Rey Juan Carlos. Calle Tulip\'{a}n S/N. 28933--M\'{o}stoles--Madrid. Spain.} 
\end{center}

\mbox{}

\mbox{}

\mbox{}

\mbox{}

\begin{center} 
{\bf Abstract}
\end{center}
\noindent

A new $n$-dimensional family of Poisson structures is globally characterized and analyzed, including the construction of its main features: the symplectic structure and the reduction to the Darboux canonical form. Examples are given that include the generalization of previously known solution families such as the separable Poisson structures.

\mbox{}

\mbox{}

\mbox{}

\mbox{}



\mbox{}


\vfill

\noindent $^1$ Telephone: (+34) 91 488 73 91. Fax: (+34) 91 664 74 55. \newline 
\mbox{} \hspace{0.05cm} E-mail: benito.hernandez@urjc.es 

\pagebreak
\begin{flushleft}
{\bf 1. Introduction}
\end{flushleft}

The presence of finite-dimensional Poisson systems in most fields of nonlinear dynamics is widespread (for instance, see \cite{olv1} for an overview and a historical discussion). In fact, very diverse physical systems have been reported to be of the Poisson kind (a sample is given in \cite{tur}-\cite{nut2} and references therein) in spite that such identification often constitutes a nontrivial issue \cite{tur,gyn1,bhqpa,prlk}. Thus, the existence of 
finite-dimensional Poisson systems of applied interest includes domains such as mechanics, electromagnetism, optics, control theory, fluid mechanics, plasma physics, population dynamics, etc. Recasting a given vector field as a Poisson system (when possible) allows the use of very diverse techniques and specific methods adapted to such format, including stability analysis, numerical integration, perturbation methods, bifurcation analysis and characterization of chaotic behavior, or determination of integability properties and invariants, just to mention a sample. For instance, see the discussions in \cite{bma1,blu1} for a brief account of such application domains and specific methods.

In terms of coordinates $x_1, \ldots , x_n$, a finite-dimensional Poisson system is a smooth dynamical system defined in a domain $\Omega \subseteq \mathbb{R}^n$ that can be expressed in the form
\begin{equation}
\label{nham}
    \dot{x} \equiv \frac{\mbox{d}x}{\mbox{d}t}= {\cal J}(x) \cdot \nabla H(x) 
\end{equation} 
where $x = (x_1, \ldots ,x_n)^T$ and superscript $^T$ denotes the transpose of a matrix. Function $H(x)$ is by construction a time-independent first integral (the Hamiltonian), and the $n \times n$ structure matrix ${\cal J}(x)$ is composed by the structure functions $J_{ij}(x)$ which are skew-symmetric
\begin{equation}
\label{sksym}
	J_{ij}(x) =  - J_{ji}(x) \:\; , \;\:\;\: i,j=1, \ldots ,n
\end{equation}
and must be also solutions of the Jacobi PDEs: 
\begin{equation}
\label{jac}
     \sum_{l=1}^n 
	\left( \begin{array}{c} J_{li}(x) \partial_l J_{jk}(x) + J_{lj}(x) \partial_l J_{ki}(x) + 
     J_{lk}(x) \partial_l J_{ij}(x) \end{array} \right) = 0 
	\:\; , \;\:\;\: i,j,k=1, \ldots ,n
\end{equation}

In addition to the diversity of specific methods and application domains already mentioned, finite-dimensional Poisson systems are of interest because they provide a broad generalization of classical Hamiltonian systems, allowing for odd-dimensional vector fields and for structure matrices much more general than the classical (constant) symplectic matrices. On the other hand, Poisson systems maintain a dynamical equivalence to classical Hamiltonian systems, at least locally, as stated by Darboux' theorem \cite{olv1}. The possible rank degeneracy of the structure matrix ${\cal J}(x)$ implies that a certain class of first integrals ($D(x)$ in what follows) termed Casimir invariants exist. There is no analog in the framework of classical Hamiltonian systems for such constants of motion, which are characterized as the solution set of the system of coupled PDEs: ${\cal J}(x) \cdot \nabla D(x) =0$. The determination of Casimir invariants and their use in order to carry out a reduction (local, in principle) is the cornerstone of the (at least local) dynamical equivalence between Poisson systems and classical Hamiltonian systems, as stated by Darboux' theorem: according to it, the level surfaces of a complete set of Casimir invariants are even-dimensional manifolds on which the Poisson system can be reduced to Hamiltonian form, at least in the neighborhood of every point. This justifies that Poisson systems can be regarded, to a large extent, as a natural generalization of classical Hamiltonian systems. However, the achievement of the Darboux reduction may be a complicated task in general, specially when it can be carried out globally. 
In fact, the global Darboux reduction is known only for a limited number of Poisson structures and systems \cite{olv1,dhns,byv2,bs2,bma1},\cite{byv4}-\cite{bn5}.

Expressing a given vector field not explicitly written in the form (\ref{nham}) in terms of a Poisson system is a nontrivial decomposition to which important efforts have been devoted in the past years in a variety of approaches \cite{tur}-\cite{byv2},\cite{nut2}-\cite{prlk}. Clearly, the source of the difficulty is twofold: First, a known constant of motion of the system able to play the role of the Hamiltonian is required. And second, it is necessary to find a suitable structure matrix for the vector field. Consequently, finding a solution of the Jacobi identities (\ref{jac}) complying also with conditions (\ref{sksym}) is unavoidable. This explains the attention deserved in the literature by the obtainment and classification of skew-symmetric solutions of the Jacobi equations \cite{olv1,tur,gyn1,bs2},\cite{bhqpa}-\cite{bn5}.
As far as the Jacobi identities constitute a set of nonlinear coupled PDEs, the characterization of globally defined solutions has followed, roughly speaking, a sequence of increasing nonlinearity and increasing dimension (for instance, see \cite{bn1,bn2} for a brief description of different solution families from the point of view of dimension and nonlinearity). In particular, there is a clear lack of knowledge of solutions verifying the following properties: (i) to be defined for arbitrary dimension $n$; (ii) for every $n$, to allow all possible (even) values of the rank; (iii) to be defined in terms of smooth functions of arbitrary degree of nonlinearity; (iv) a complete set of functionally independent Casimir invariants can be explicitly determined; (v) it is also possible to construct the Darboux canonical form; (vi) the previous items can be achieved globally in phase space. The number of known solution families for which all these general conditions hold simultaneously is very limited \cite{olv1,byv4,bn3}. 

In this work, a new family of skew-symmetric solutions of the Jacobi equations is characterized and analyzed. Such family generalizes the separable structure matrices \cite{byv4}, a well-known family presenting the remarkable feature of complying to all the conditions (i)-(vi) just enumerated. Moreover, the new solution family reported in the present work also verifies the six conditions with full generality. Accordingly, such new family of Poisson structures is not only characterized but its global analysis is also carried out in what follows. In addition, it can be seen that previously known types of Poisson structures appearing in a diversity of physical situations now become particular cases of the new solution family, as it will be illustrated in the examples section. 

The structure of the article is the following. In Section 2 the new family of global Poisson structures is characterized and its global analysis (including the reduction to the Darboux canonical form) is constructively developed. Section 3 is devoted to the presentation of examples, one of them showing the generalization of the family of separable structure matrices. The work finishes in Section 4 with some remarks on the method and possible future outlines. 

\mbox{}

\begin{flushleft}
{\bf 2. New family of solutions and its global analysis}
\end{flushleft}

The fundamental result of this article is presented in this section. As indicated in the Introduction, $x = (x_1, \ldots ,x_n)^T$. Similar notation shall be employed for the $n$-d vectors of variables $y = (y_1, \ldots ,y_n)^T$ and $z = (z_1, \ldots ,z_n)^T$. Additionally, in what follows the character $w$ shall denote a single real variable. Also, for every 
one-variable real function $f(w)$ its derivative will be denoted $f'(w)$ and its inverse function $f^{-1}(w)$, namely $f(f^{-1}(w))=f^{-1}(f(w))=w$.

\mbox{}

\noindent{\bf Theorem 1.} 
{\em Let $S \equiv (S_{ij})$ be an $n \times n$ skew-symmetric real matrix of rank $r \leq n$, let $L \equiv (L_{ij})$ be an $n \times n$ invertible real matrix, and let $\Lambda \equiv (\Lambda_{ij})$ be the inverse of $L$. For $i=1, \ldots ,n$, let $\Omega_i \subseteq \mathbb{R}$ be real domains, and define $\Omega \subseteq \mathbb{R}^n$ to be the domain $\Omega \equiv \Omega_1 \times \ldots \times \Omega_n$. In addition, for every $i=1, \ldots ,n$, let $\psi_i(w) : \Omega_i \rightarrow \mathbb{R} - \{ 0 \}$ be a one-variable real function which is $C^{\infty}(\Omega_i)$ and nonvanishing in $\Omega_i$. Similarly, for every $i=1, \ldots ,n$, let $\lambda_i(x) : \Omega \rightarrow \Omega_i$ be the real function defined as $\lambda_i(x) \equiv \sum_{j=1}^n \Lambda_{ij} x_j$. We then have:
\begin{description}
\item a) The following set of functions 
\begin{equation}
\label{sgs}
	J_{ij}(x) = \sum_{k,l=1}^n L_{ik}S_{kl}L_{jl} 
			\psi_k \left( \lambda_k(x) \right) \psi_l \left( \lambda_l(x) \right) 
			\;\:\: , \:\;\:\; i,j=1, \ldots ,n
\end{equation}
constitute the entries of an $n \times n$ structure matrix ${\cal J}(x) \equiv (J_{ij}(x))$ globally defined in $\Omega$ and of rank $r$ everywhere in $\Omega$. 
\item b) Let $\{ k^{[r+1]}, \ldots ,k^{[n]} \}$ be a basis of Ker$\{ S \}$, where $k^{[i]}=(k^{[i]}_1, \ldots ,k^{[i]}_n)^T$ for $i=r+1, \ldots ,n$. In addition, consider the functions 
\begin{equation}
\label{xici}
	\xi_i(w) = \int \frac{\mbox{\rm d}w}{\psi_i(w)} \;\:\: , \:\;\:\; i=1, \ldots ,n
\end{equation}
which are primitives defined in $\Omega_i$ for which arbitrary integration constants can be chosen. Then for every structure matrix ${\cal J}(x)$ of the form (\ref{sgs}) a complete set of Casimir invariants globally defined and functionally independent in $\Omega$ is given by:
\begin{equation}
\label{gci}
    D_i(x) = \sum _{j=1}^{n} k^{[i]}_j \xi_j \left( \lambda_j(x) \right) 
	    \;\:\: , \:\;\:\; i = r+1, \ldots, n
\end{equation}
\item c) Every structure matrix ${\cal J}(x)$ of the form (\ref{sgs}) can be constructively and globally reduced in $\Omega$ to the Darboux canonical form. 
\end{description}
}

\mbox{}

\noindent{\bf Proof.} 
The proof is constructive. In first place, let us show that there exists a set of $n$ smooth and invertible functions $\phi_i(w)$ for all $i=1, \ldots ,n$, such that:
\begin{equation}
\label{fimu}
	\phi_i^{-1}(w) = \xi_i(w) = \int \frac{\mbox{d}w}{\psi_i(w)} 
	\;\:\; \mbox{\rm in} \;\: \Omega_i   \;\:\: , \;\:\;\: i = 1, \ldots ,n 
\end{equation}
In (\ref{fimu}) it is assumed that a given (arbitrary) choice of the integration constant is made for every function $\phi_i^{-1}(w)$. Now for all $i=1, \ldots ,n$, it is $\psi_i(w) \neq 0$ in $\Omega_i$, and thus we have $\left( \phi_i^{-1}(w) \right)' = 1/ \psi_i(w) \neq 0$. Therefore $\phi_i (w)$ exists, and it is unique, smooth and invertible in the domain $\Omega^*_i=\phi_i^{-1}( \Omega_i)$ for every $i$. Then, using the chain rule for the inverse function we find:
\begin{equation}
\label{psimu}
	\psi_i(w) = \frac{1}{\left( \phi_i^{-1}(w) \right)'} = 
	\phi_i' \left( \phi_i^{-1}(w) \right) \;\: , \;\:\;\: i = 1, \ldots ,n
\end{equation}
Taking these elements into account, we now define the following smooth diffeomorphic transformation globally defined in $\Omega$: 
\begin{equation}
\label{dtr1}
	y_i(x) = \phi_i^{-1}( \lambda_i(x)) \;\: , \;\:\;\: i = 1, \ldots ,n
\end{equation}
Let us now recall the transformation rule of a structure matrix ${\cal J}(x) \equiv (J_{ij}(x))$ after a smooth diffeomorphism $y \equiv y(x)$. The result is a new structure matrix ${\cal J}^*(y) \equiv (J^*_{ij}(y))$ defined in $y( \Omega)$:
\begin{equation}
\label{jdiff}
      J^*_{ij}(y) = \sum_{k,l=1}^n \frac{\partial y_i}{\partial x_k} J_{kl}(x) 
	\frac{\partial y_j}{\partial x_l} \;\: , \;\:\;\: i,j = 1, \ldots ,n
\end{equation}
Now let us apply identities (\ref{jdiff}) for transformation (\ref{dtr1}) to matrix 
(\ref{sgs}). We can rewrite (\ref{sgs}) as: 
\[
	J_{ij}(x(y)) = \sum_{k,l=1}^n L_{ik}S_{kl}L_{jl} \phi_k'(y_k) \phi_l'(y_l)
\]
In addition, making use again of the chain rule (\ref{psimu}) for the inverse function we have:
\[
	\frac{\partial y_i}{\partial x_j} = 
	\Lambda_{ij} \left( \phi_i^{-1} \right) ' (\phi_i(y_i)) = 
	\frac{\Lambda_{ij}}{\phi_i'(y_i)}
\]
The outcome is then:
\[
	J^*_{ij}(y) = \sum_{k,l,r,s=1}^n \Lambda_{ik} \Lambda_{jl} L_{kr} S_{rs} L_{ls} 
	\frac{1}{\phi_i'(y_i)} \frac{1}{\phi_j'(y_j)} \phi_r'(y_r) \phi_s'(y_s) 
\]
After some algebra, this reduces to $J^*_{ij}(y) = S_{ij}$, namely ${\cal J}^*(y)=S$. Since $S$ is constant and skew-symmetric, it is in fact a structure matrix defined in $\mathbb{R}^n$. As a consequence of diffeomorphism (\ref{dtr1}), this implies that ${\cal J}(x)$ is also a structure matrix in $\Omega$. In addition, it is by construction $\mbox{\rm Rank} \{ {\cal J}(x) \} = \mbox{\rm Rank} \{ S \} =r$ everywhere in $\Omega$. This proves statement (a) of Theorem 1. 

Let us now consider statement (b). We take as starting point the structure matrix ${\cal J}^*(y)=S$ obtained from ${\cal J}(x)$ after transformation (\ref{dtr1}). Let $\{ k^{[r+1]}, \ldots ,k^{[n]} \}$ be a basis of Ker$\{ S \}$, where $k^{[i]}=(k^{[i]}_1, \ldots ,k^{[i]}_n)^T$ for $i=r+1, \ldots ,n$. According to \cite{byv4}, a complete set of Casimir invariants globally defined and functionally independent in ${\mathbb R}^n$ for the structure matrix $S$ is given by: 
\begin{equation}
\label{gctyy}
    D^*_i(y) = \sum _{j=1}^{n} k^{[i]}_j y_j \;\:\: , \:\;\:\; i = r+1, \ldots, n
\end{equation}
Recall that the set of Casimir functions (\ref{gctyy}) is mapped into a complete set of globally defined and functionally independent Casimir invariants $D_i(x)$ of ${\cal J}(x)$ in $\Omega$ according to the transformation rule: $D_i(x) = D^*_i \left( y(x) \right)$ for $i = r+1, \ldots, n$. Taking into account (\ref{fimu}) and (\ref{dtr1}), substitution in 
(\ref{gctyy}) leads to expression (\ref{gci}) and statement (b) of Theorem 1 is proved.

We proceed now to show statement (c). We again take as starting point matrix ${\cal J}^*(y)=S$ previously constructed. A second transformation which is also globally defined in $\mathbb{R}^n$ is performed:
\begin{equation}
\label{trp}
   z_i = \sum _{j=1}^n P_{ij} y_j \:\; , \:\;\:\; i = 1, \ldots , n
\end{equation}
where $P \equiv (P_{ij})$ is a constant, $n \times n$ invertible matrix. According to 
(\ref{jdiff}) and (\ref{trp}), the structure matrix ${\cal J}^*(y)=S$ is transformed into a new structure matrix ${\cal J}^{**}(z) \equiv (J^{**}_{ij}(z))$ given by:
\begin{equation}
   \label{trp2}
   {\cal J}^{**}(z) = P \cdot {\cal J}^* (y) \cdot P^T = P \cdot S \cdot P^T
\end{equation}
A classical result of linear algebra \cite{ayr1} shows that matrix $P$ in (\ref{trp2}) can always be chosen in such a way that:
\begin{equation}
\label{jhamth1}
	{\cal J}^{**}(z) = \left( \begin{array}{cc} 0 & 1 \\ -1 & 0 \end{array} \right) 
	\overbrace{ \oplus \ldots \oplus }^{(r/2)} 
	\left( \begin{array}{cc} 0 & 1 \\ -1 & 0 \end{array} \right) \oplus O_1 
	\overbrace{ \oplus \ldots \oplus }^{(n-r)} O_1
\end{equation}
where $r = \mbox{\rm Rank} \{ S \}$ is even because $S$ is skew-symmetric, and $O_1 \equiv (0)$ denotes the $1 \times 1$ null matrix. Therefore, in (\ref{jhamth1}) the original structure matrix ${\cal J}(x)$ has been reduced globally to the Darboux canonical form, since ${\cal J}^{**}(z)$ is the direct sum of $\, (r/2) \,$ symplectic $2 \times 2$ matrices plus $\, (n-r) \,$ null $1 \times 1$ matrices $O_1$ associated with the Casimir invariants, which in the Darboux representation are decoupled and correspond in (\ref{jhamth1}) to the variables $z_{r+1}, \ldots , z_n$. This shows statement (c) and completes the proof. 
$\:\;\:\; \Box$

\mbox{}

The main presentation of results is thus concluded. In the next section, some applied examples are analyzed. 

\mbox{}

\begin{flushleft}
{\bf 3. Examples}
\end{flushleft}

\noindent {\bf Example 1.} {\em Separable solutions.}

\mbox{}

Consider the solutions described in Theorem 1 in the particular case in which matrix $L$ is the $n \times n$ identity matrix. Taking into account that $L_{ij}= \Lambda _{ij}= \delta _{ij}$ for all $i,j=1, \ldots ,n$, we see that now it is $\lambda_i(x)=x_i$ for all $i$ and thus, after some direct calculations it is found that solutions (\ref{sgs}) are reduced to: $J_{ij}(x)=S_{ij} \psi_i(x_i) \psi_j(x_j)$, for $i,j=1, \ldots ,n$. These are $n$-dimensional and globally defined structure matrices of arbitrary rank and degree of nonlinearity. They are termed {\em separable\/} after their functional form resembling the classical separation of variables technique for PDEs. The analytic characterization of separable structure matrices, as well as their symplectic structure and the global reduction to the Darboux canonical form were presented in \cite{byv4}. Now all these features appear as particular cases of Theorem 1. For instance, according to (\ref{xici}-\ref{gci}) we find that a complete set of independent Casimir invariants is given by
\[
    D_i(x) = \sum _{j=1}^{n} k^{[i]}_j \int \frac{\mbox{\rm d}x_j}{\psi_j(x_j)}
	    \;\:\: , \:\;\:\; i = r+1, \ldots, n
\]
which is the result found in \cite{byv4}. Similarly, for the reduction of separable structure matrices to the Darboux canonical form we can apply the general procedure developed in the proof of Theorem 1. According to such proof, we see from equation (\ref{dtr1}) that the first transformation to be performed becomes in the separable case:
\begin{equation}
\label{des}
	y_i(x) = \phi_i^{-1}(x_i) = \int \frac{\mbox{\rm d}x_i}{\psi_i(x_i)}
	    \;\:\: , \:\;\:\; i = 1, \ldots, n
\end{equation}
Again, diffeomorphism (\ref{des}) is the one characterized in \cite{byv4}. Taking the transformation rule (\ref{jdiff}) into account we arrive to ${\cal J}^*(y)=S$. Then, after this point the reduction to the Darboux canonical form carried out in \cite{byv4} exactly follows the general procedure developed in the proof of Theorem 1. To complete these considerations, it is worth recalling that separable structure matrices embrace a wide variety of physical systems of interest in the literature, such as Poisson realizations of Lotka-Volterra and Generalized Lotka-Volterra systems (including examples of relevance in both plasma physics and population dynamics), diverse instances of Toda lattices and relativistic Toda lattices, all constant structure matrices (of which the entire classical Hamiltonian theory is a particular case), the separable formulation of the Kermack-McKendrick model (see also Example 2), a differential model associated with circle maps, and Poisson systems arising in the study of $2 \times 2$ games (see \cite{byv4} for full details regarding all these physical instances of separable solutions). Consequently, separable structure matrices constitute a relevant family of solutions which is now generalized by Poisson structure matrices (\ref{sgs}) characterized in Theorem 1.  

\mbox{}

\noindent {\bf Example 2.} {\em A nonseparable instance: Kermack-McKendrick model.}

\mbox{}

In order to illustrate the generality and scope of Theorem 1, it is also interesting to describe one applied instance which is not separable. For this, we shall consider the Poisson formulation of the Kermack-McKendrick model of population dynamics \cite{gyn1,nut2}. This system admits a biHamiltonian description in which one of the structure matrices is separable (omitted here for the sake of conciseness, see \cite{byv4} for details) and thus belongs to the framework considered in Example 1. The second structure matrix is not separable, and is the one of interest in what follows: 
\begin{equation}
\label{kmk}
	{\cal J}(x) = rx_1x_2 \left( \begin{array}{ccc} 0 & 1 & -1 \\ -1 & 0 & 1 \\ 1 & -1 & 0 
	\end{array} \right)
\end{equation}
where $r>0$ is a real constant. Since the variables describe biological populations, we have $x_i>0$, namely $\Omega_i= \mathbb{R}^+ - \{ 0 \}$ for all $i=1,2,3,$ and $\Omega = \Omega_1 \times \Omega_2 \times \Omega_3$ is the positive 3-d orthant. Thus we have Rank$\{ {\cal J} \} =2$ in $\Omega$. After some calculations it can be verified that structure matrix (\ref{kmk}) is described in terms of Theorem 1 with $\psi_1(w)= \psi_2(w)=\sqrt{r}w$, $\psi_{3}(w)=1$, and matrices: 
\[
	S = \left( \begin{array}{ccc} 0 & 1 & 0 \\ -1 & 0 & 0 \\ 0 & 0 & 0 \end{array} \right) 
	\:\; , \:\;\:\:\;
	L = \left( \begin{array}{ccc} 1 & 0 & 0 \\ 0 & 1 & 0 \\ -1 & -1 & 1 \end{array} \right)
	\:\; , \:\;\:\:\;
	\Lambda = L^{-1} = 
		\left( \begin{array}{ccc} 1 & 0 & 0 \\ 0 & 1 & 0 \\ 1 & 1 & 1 \end{array} \right)
\]
It is worth noting that in this case it is $\lambda_1(x)=x_1$, $\lambda_2(x)=x_2$ and $\lambda_3(x)=x_1+x_2+x_3$. Accordingly, it is $\psi_1(\lambda_1(x))=\sqrt{r}x_1$, $\psi_2(\lambda_2(x))=\sqrt{r}x_2$ and $\psi_3(\lambda_3(x))=1$. From this, the construction of the structure matrix (\ref{kmk}) is straightforward in terms of the elements indicated in Theorem 1. In addition, use of (\ref{gci}) can be made in order to find the only independent Casimir invariant. For this, we have that a basis of Ker$\{ S \}$ is given by $\{ k^{[3]} \} = \{ (0,0,1)^T \}$, and $\xi_3(w)=w$. Consequently, it is $D(x) = \xi_3( \lambda_3(x)) = x_1+x_2+x_3$. In addition, for the global reduction to the Darboux canonical form the construction developed in Theorem 1 is again followed. According to (\ref{dtr1}) the first diffeomorphic transformation to be performed is given by $y_i(x) = \phi_i^{-1}( \lambda_i(x))$ for $i=1,2,3$. Following the definition of the $\phi_i(w)$ given in (\ref{fimu}), we have $\phi_1(w)= \phi_2(w)= \exp (\sqrt{r}w)$ and $\phi_3(w)=w$. 
Consequently, now the diffeomorphic transformation (\ref{dtr1}) in $\Omega$ amounts to:
\[
	y_1(x) = \frac{1}{\sqrt{r}} \ln (x_1) \:\; , \:\;\:\, 
	y_2(x) = \frac{1}{\sqrt{r}} \ln (x_2) \:\; , \:\;\:\, 
	y_3(x) = x_1+x_2+x_3
\]
After some algebra, and taking relationship (\ref{jdiff}) into account, the transformed structure matrix is ${\cal J}^*(y)=S$, which actually corresponds to the Darboux canonical form (\ref{jhamth1}). We thus see that this nonseparable Poisson structure is also described in the framework of Theorem 1. 

\mbox{}

\begin{flushleft}
{\bf 4. Final remarks}
\end{flushleft}

In this contribution a new family of Poisson structures has been characterized. Such family 
comprises structure matrices of arbitrary dimension and rank, can be globally defined and analyzed, consists of functions of an arbitrary degree of nonlinearity, and is such that a complete set of globally defined independent Casimir invariants can be characterized. Moreover, such solution family can be globally and constructively reduced to the Darboux canonical form. As indicated in the Introduction, the number of known families of Poisson structures satisfying all such properties is very limited. In addition, one of those known families (the separable structure matrices \cite{byv4}) is generalized by the family characterized in Theorem 1. In fact, such generalization is proper (namely the solution family reported in Theorem 1 is not limited to separable solutions, as shown in Example 2). Accordingly, a significant number of physically relevant Poisson structures and systems becomes unified in the common framework provided by Theorem 1. Therefore these systems need not to be considered separately by {\em ad hoc\/} approaches, but inspected in a more general and systematic frame. All these results show that the investigation of Poisson structures and systems from the point of view of the Jacobi equations provides a fruitful perspective that deserves further consideration. 

\pagebreak

\end{document}